# Wakefield-induced THz wave generation in a hybrid dielectric-plasma cylindrical waveguide

A. A. Molavi Choobini and F. M. Aghamir*

Dept. of Physics, University of Tehran, Tehran 14399-55961, Iran.

**Abstract:**

In the present study, the generation of THz radiation through wakefield excitation in a cylindrical dielectric plasma waveguide is investigated. The proposed hybrid dielectric-plasma wakefield structure combines the advantages of dielectric materials and plasma, creating a versatile platform for high-performance applications and advanced THz radiation generation. By leveraging the strengths of both techniques, this hybrid configuration achieves enhanced gradients and optimized THz wave output. The mechanism involves a high-energy laser pulse propagating through a plasma-loaded dielectric waveguide, inducing wakefields that drive THz wave emission. This dual capability underscores the versatility of the proposed structure, offering significant advancements in THz wave generation technologies. To support the theoretical analysis, numerical simulations were employed using the Fourier-Bessel Particle-In-Cell (FBPIC) method and the COMSOL Multiphysics package. The simulation modeled the wakefield generation process and validated the propagation of electromagnetic waves. A comprehensive parametric study examined the effects of various parameters including dielectric thickness, an external DC magnetic field, laser pulse length, driver beam radius, and total current, on the wakefield generated THz radiation. Through systematic variation of these parameters, the study aims to elucidate the controlled features of the resulting fields and optimize THz radiation.

## I. Introduction

The generation of terahertz (THz) waves exhibits several distinctive characteristics, such as non-ionizing interactions with matter, high spatial resolution, and the ability to penetrate various materials [1-4]. These properties, coupled with their wide-ranging applications in imaging, spectroscopy, communication, and sensing technologies, have made THz waves a key focus of recent scientific investigation [5, 6]. However, the development of efficient and compact THz radiation sources remains a significant challenge, driving exploration into innovative generation mechanisms [7, 8]. Recent advances in laser and beam technologies have positioned wakefield-based methods as a particularly promising approach, attracting substantial interest [9–11]. Wakefield generation exploits the intense electromagnetic fields created when a high-energy particle beam or ultra-intense laser pulse propagates through a medium. As the driving beam or pulse passes through, it perturbs the medium's equilibrium, producing oscillating electromagnetic fields—known as wakefields—that trail the driver. These wakefields, with their distinctive spatial structure and dynamic behavior, enable efficient energy transfer to charged particles and facilitate the generation of high-frequency radiation, including THz waves [12- 16].

R. R. Knyazev and colleagues examined the effect of plasma on the amplitude of wakefields excited by a relativistic electron bunch in a dielectric structure [17]. Their research showed that plasma waves influence the peak field by tuning the eigenwave frequencies to match the bunch repetition frequency through adjustments to structural radii. B. D. O'Shea and colleagues investigated THz radiation generation using relativistic electron beams in a dielectric-lined waveguide [18]. They observed a reversible, field-induced damping effect at THz frequencies, attributed to changes in material conductivity, achieving field strengths exceeding 2GV/m for narrowband THz sources. L. Shi et al. proposed a compact, high-power THz source that combines laser plasma wakefield acceleration and dielectric-lined waveguides [19]. This method generates mJ-scale narrowband THz pulses suitable for driving nonlinear phenomena in condensed matter. M. I. Ivanyan and colleagues analyzed a two-layer circular metal-dielectric waveguide, accounting for energy dissipation within the dielectric layer due to radiation [20]. They evaluated the longitudinal wake potentials for thin and thick internal layers, revealing that the waveguide supports only a slowly propagating TM mode. Hanqi Feng et al. introduced a method for generating tunable, narrowband THz pulses with 10-mJ-level energy using high-density electron bunch trains modulated by nonlinear plasma wakefields [21]. This approach achieves efficient microbunching with high charge and bunching factors, enabling THz radiation for high-repetition-rate accelerator facilities. Mitchell E. Schneider and co-workers proposed a compact, tunable THz source powered by an electron beam passing through a dielectric-loaded traveling wave structure [22]. Their scheme achieves high peak power THz generation in 0.4–1.6 THz range with 6.8% efficiency, offering portability for various applications. G. Lehmann and K. H. Spatschek investigated THz radiation generation from a plasma grating formed by two counter-propagating long laser pulses [23]. Using PIC simulations, they identify electromagnetic wave emission near the plasma frequency, driven by instability between the wakefield and plasma grating transitioning to THz radiation at the grating boundaries. Rezaei-Pandari and colleagues explored THz radiation generation from laser-wakefield in a helium gas using 3D particle-in-cell simulations [24]. Their analysis of electromagnetic fields, emission angles, and spectra, research demonstrated laser-wakefield efficiency as a THz source and highlighted the benefits of a moving simulation window for accurate THz studies.

The present study introduces a hybrid dielectric-plasma wake-field cylindrical structure, leveraging the strengths of both dielectric materials and plasmas to create a versatile platform for advanced THz radiation generation and high-performance applications. This approach combines the strong electromagnetic field confinement of dielectric materials with the unique ability properties of plasmas, the ability to sustain ultra-high electric fields without electrical breakdown. This dual advantage ensures that a larger fraction of the input laser energy is converted into THz radiation, making the process more energy-efficient than purely dielectric or plasma-based systems. By integrating these complementary properties, the proposed structure overcomes the limitations of conventional THz sources, such as challenges in simultaneously achieving high power, broad bandwidth, and coherence. These limitations often restrict the effectiveness of THz waves for applications in spectroscopy, imaging, and communications. The dielectric-plasma wake-field structure employs a high-energy laser pulse propagating through a dielectric medium embedded with plasma, facilitating efficient wake-field excitation in the THz frequency range. The laser pulse, perturbs the plasma electrons, inducing oscillatory motion that generates coherent, high-power THz radiation. This innovative approach underscores the transformative potential of the hybrid dielectric-plasma wakefields in advancing THz wave technologies. The article is structured into the following segments: Section II examines the theoretical model of the dielectric-plasma wake-field wave generator. Section III discusses the outcome and characteristics of wakefields. Conclusions are drawn in section IV.

## II. Theoretical Model

The configuration of wakefield-induced THz wave generation in a hybrid dielectric-plasma cylindrical waveguide is considered, as depicted in Fig. 1. The dielectric-plasma cylindrical structure is assumed to extend infinitely along the z-axis. The system is driven by an azimuthally symmetric charge and current density source, expressed as:

$$\rho = \rho_0 \delta(z - vt)\Theta(r - r_d) \quad (1)$$

and

$$\vec{J} = \rho_0 v_0 \delta(z - vt)\Theta(r - r_d)\hat{e}_k \quad (2)$$

where $r_d$ is the driver beam radius, $\rho_0$ is the beam charge density, and $\Theta$ is the Heaviside step function. Given the axial symmetry of the charge and current distributions and the absence of a radial current source, the analysis exclusively focuses on azimuthally symmetric modes with longitudinal electric fields ($TM_{0n}$ modes). TEM modes are absent in a hollow, perfectly symmetric metal conductor without radial current components. These modes require both transverse electric and magnetic fields with no longitudinal components, which are unsustainable without a radial current source.

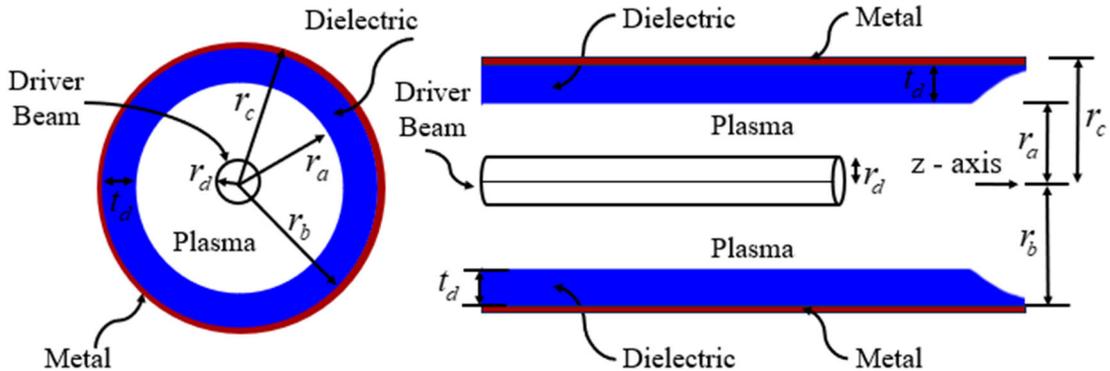

**Fig. 1.** Schematic of dielectric-plasma waveguide structure and cross-section for wakefield wave generation.

The driven beam excites transverse electromagnetic modes in this wakefield-induced THz structure due to the Cherenkov effect. The radiation occurs because the drive beam travels at velocity $v = c/\sqrt{\epsilon}$ through the dielectric and plasma, where $\epsilon$ is the permittivity of the dielectric or plasma, and $c$ is the speed of light. The phase velocity of the generated wakefield wave matches the driving beam's velocity. For electrons moving with the velocity $v$ along the structure axis (z-direction), the wakefield wave in the dielectric-plasma structure depends on the amplitude and frequency of each mode. The structure is assumed to be sufficiently long and ensure that the wakefield modes are fully developed along the z-axis. The significant impedance contrast between the dielectric and plasma materials inherently guides and confines the electromagnetic fields at the dielectric-plasma interface. Furthermore, the axial symmetry of the charge and current distributions enhances the stability of the $TM_{0n}$ modes, minimizing the likelihood of field leakage or instability. To derive the wave equation for the $n$-th axial component of the TM mode, the linearized fluid and the wave equation are derived from Maxwell's equations through mathematical manipulation as follows:

$$\nabla^2 \vec{E} - \frac{\epsilon}{c^2}\frac{\partial^2 \vec{E}}{\partial t^2} = \frac{4\pi}{c^2}\frac{\partial \vec{J}}{\partial t} + 4\pi\vec{\nabla}\rho \tag{3}$$

The axial symmetry of the charge and current distributions, along with the absence of a radial current source, is considered. Furthermore, the axial electric field $E_z$ of the TM mode is assumed to have a time-harmonic dependence, expressed as:

$$E_{zn}(r,z,t) = \frac{1}{2\pi v}\int_{-\infty}^{+\infty} E_z(r,\omega)e^{i\frac{\omega}{v}(z-vt)}d\omega \tag{4}$$

Upon the application of the Fourier transform to the wave equation the representation of axial field solutions within the regions of the dielectric-plasma structure are as follows:

$$\left[\frac{1}{r}\frac{\partial}{\partial r}r\frac{\partial}{\partial r} - \frac{\omega^2}{c^2\gamma^2}\right]E_{zn}^{(1)}(r,\omega) = \frac{4\pi i \rho_0 \omega}{v\gamma^2} \qquad r \leq r_d \tag{5a}$$

$$\left[\frac{1}{r}\frac{\partial}{\partial r}r\frac{\partial}{\partial r} - \frac{\omega^2}{c^2\gamma^2}\right]E_{zn}^{(2)}(r,\omega) = 0 \qquad r_d < r < r_a \tag{5b}$$

$$\left[\frac{1}{r}\frac{\partial}{\partial r}r\frac{\partial}{\partial r} - \frac{\omega^2}{c^2}\left(1-\epsilon\frac{v^2}{c^2}\right)\right]E_{zn}^{(3)}(r,\omega) = 0 \qquad r_a < r < r_b \tag{5c}$$

where $r_a$ and $r_b$ are the inner and outer radii of the dielectric-plasma liner, $\gamma^2 = \left(1-\frac{v^2}{c^2}\right)^{-1}$ is the relativistic factor, and the following relationship is used for the delta function along the z-axis,

$$\delta(z-vt) = \frac{1}{2\pi v}\int_{-\infty}^{+\infty} e^{i\frac{\omega}{v}(z-vt)}d\omega \tag{6}$$

The axial electric field distributions within the different regions of the dielectric-plasma structure can be expressed as:

$$E_{zn}^{(1)}(r,\omega) = -\frac{4\pi i \rho_0 \omega}{v} + A_1 I_0\left(\frac{\omega\gamma}{v}r\right) \qquad r \leq r_d \tag{7a}$$

$$E_{zn}^{(2)}(r,\omega) = A_2 I_0\left(\frac{\omega\gamma}{v}r\right) + A_3 K_0\left(\frac{\omega\gamma}{v}r\right) \qquad r_d < r < r_a \tag{7b}$$

$$E_{zn}^{(3)}(r,\omega) = A_4 J_0(\kappa r) + A_5 Y_0(\kappa r) \qquad r_a < r < r_b \qquad (7c)$$

here $\kappa^2 = \frac{\omega^2}{v^2}(\epsilon v^2/c^2 - 1)$, $J_0(z)$ is the Bessel function of the first kind, $I_0(z)$ and $K_0(z)$ are modified Bessel functions of the first and second kind, respectively. The solution for $E_{zn}^{(3)}$ satisfies the boundary condition $E_{zn}^{(3)}(r = r_b) = 0$. To determine the wakefield amplitude, appropriate boundary conditions must be applied. Imposing boundary conditions at the beam-plasma and dielectric-plasma interfaces, the axial wakefield amplitude within the beam region can be expressed as:

$$E_{zn}^{(1)}(r,\omega) = -4\pi i \rho_0 \left\{ \frac{v}{\omega}\left[1 - \frac{\omega r_d}{v\gamma}K_1\left(\frac{\omega\gamma}{v}r_d\right)\right] + \frac{r_d\left(\frac{\epsilon v^2}{c^2}-1\right)^{\frac{1}{2}} I_1\left(\frac{\omega\gamma}{v}r_d\right) K_1\left(\frac{\omega\gamma}{v}r_a\right) F_1(\kappa r_a,\kappa r_b) + \frac{\epsilon}{\gamma}G_1(\kappa r_a,\kappa r_b)K_0\left(\frac{\omega\gamma}{v}r_a\right)}{\gamma\left[I_1\left(\frac{\omega\gamma}{v}r_a\right)\left(\frac{\epsilon v^2}{c^2}-1\right)^{\frac{1}{2}} F_1(\kappa r_a,\kappa r_b) - \frac{\epsilon}{\gamma}G_1(\kappa r_a,\kappa r_b)I_0\left(\frac{\omega\gamma}{v}r_a\right)\right]} \right\} I_0\left(\frac{\omega\gamma}{v}r\right) \qquad (8)$$

where the $F_1(\kappa r_a, \kappa r_b)$ and $G_1(\kappa r_a, \kappa r_b)$ are defined as:

$$F_1(\kappa r_a, \kappa r_b) = J_0(\kappa r_a)Y_0(\kappa r_b) - J_0(\kappa r_b)Y_0(\kappa r_a) \qquad (9a)$$

$$G_1(\kappa r_a, \kappa r_b) = J_1(\kappa r_a)Y_0(\kappa r_b) - J_0(\kappa r_b)Y_1(\kappa r_a) \qquad (9b)$$

here $Y_0(z)$ and $Y_1(z)$ are the Bessel functions of the second kind of order 0 and 1, respectively. The radial electric field ($E_r$) and azimuthal magnetic field ($B_\theta$) components are evaluated through the application of the boundary conditions for the continuity of the axial electric field ($E_z$) and the radial displacement vector at the interface. The above derivation utilizes the following relations:

$$E_r = -\frac{i}{\frac{\omega}{v}(1-\epsilon v^2/c^2)}\frac{\partial E_z}{\partial r} \qquad (10a)$$

$$B_\theta = \frac{\epsilon v}{c} E_r \qquad (10b)$$

The spatial variation of the axial electric field is related to the radial and azimuthal components of the electromagnetic field (as expressed by Eqns. 10a and b), ensuring the self-consistent solutions in which all components of fields satisfy Maxwell's equations and the boundary conditions imposed by the waveguide's geometry and material properties. Given that most wakefield devices employ drive beams with $v \approx c$ and taking the limit as $v$ approaches $c$, Eq. (8) becomes:

$$E_{zn}^{(1)}(r,\omega) = -4\pi i \rho_0 \left(\frac{r_d^2}{r_a}(\epsilon-1)^{\frac{1}{2}}\right)\left(\frac{F_1(\zeta r_a,\zeta r_b)}{r_a\zeta F_1(\zeta r_a,\zeta r_b) - 2\epsilon G_1(\zeta r_a,\zeta r_b)}\right) \qquad (11)$$

where $\zeta = \frac{\omega}{v}(\epsilon-1)^{\frac{1}{2}}$. The denominator of Eq. (11) represents the dispersion relation of the dielectric-plasma structure under relativistic condition. Inverse Fourier transform of Eq. (11) and subsequent evaluation using contour integration leads to an expression for the time-varying axial wakefield:

$$E_{zn}^{(1)}(r,\xi) = \sum_{n=0}^{\infty} \frac{8I_b}{r_a c} \frac{\sqrt{\epsilon-1} F_1(\zeta r_a, \zeta r_b)}{\Psi(\Lambda_n, \epsilon, r_a, r_b)} Cos\Lambda_n \xi \qquad (12)$$

Where $\xi$ is an independent variable defined as $\xi = z - ct$, $I_b = \pi \rho_0 r_d^2 c$ is the total current of the waveguide structure, and $\Psi(\Lambda_n, \epsilon, r_a, r_b)$ is mode function:

$$\Psi(\Lambda_n, \epsilon, r_a, r_b) = \sqrt{\epsilon-1}\left[r_a(1-2\epsilon)F_1(\kappa r_a, \kappa r_b) + \left(\frac{2\epsilon}{\zeta} - \zeta r_a^2\right)G_1(\kappa r_a, \kappa r_b) + \right.$$
$$\left. \zeta r_a r_b F_2(\kappa r_a, \kappa r_b) - 2\epsilon r_b G_2(\kappa r_a, \kappa r_b)\right] \qquad (13)$$

here $F_2(\kappa r_a, \kappa r_b)$ and $G_2(\kappa r_a, \kappa r_b)$ are:

$$F_2(\zeta r_a, \zeta r_b) = Y_0(\zeta r_a)J_1(\zeta r_b) - J_0(\zeta r_a)Y_1(\zeta r_b) \qquad (14a)$$

$$G_2(\zeta r_a, \zeta r_b) = J_1(\zeta r_b)Y_1(\zeta r_a) - J_1(\zeta r_a)Y_1(\zeta r_b) \qquad (14b)$$

and $\Lambda_n = \zeta_n/\sqrt{\epsilon-1}$ where $\zeta_n$ are the roots of the following equation:

$$\zeta r_a r_b F_2(\kappa r_a, \kappa r_b) - 2\epsilon r_b G_2(\kappa r_a, \kappa r_b) = 0 \qquad (15)$$

Since shaping the drive pulse reduces the amplitude of higher-order modes within the dielectric-plasma structure and requires an asymmetric driving pulse, the current pulse distribution can be represented as:

$$\vec{J}(r,\xi) = \frac{\rho_0 c \xi}{L} \Theta(L-\xi)\Theta(r-r_d)\hat{e}_k \qquad (16)$$

Here, $L$ represents the pulse length and $\rho_0 c$ denotes the peak beam current density. Equation (12), along with the application of the convolution theorem, yields the axial wakefield generated behind the drive beam ($\xi > L$) within the dielectric-plasma structure:

$$E_{zn}^{(1)}(r,\xi) = \sum_n \frac{8I_b}{r_a L c} \frac{\sqrt{\epsilon-1}F_1(\zeta_n r_a, \zeta_n r_b)}{\Psi(\Lambda_n, \epsilon, r_a, r_b)} \left[\frac{Cos\Lambda_n(\xi-L)}{\Lambda_n^2} - \frac{Cos\Lambda_n \xi}{\Lambda_n^2} - \frac{L}{\Lambda_n}Sin\Lambda_n \xi\right] \qquad (17)$$

Eq. (17) shows a negligible radial wakefield component within the region occupied by the drive beam ($\epsilon = 1$). This suggests that the dominant wakefield effects in this region are primarily longitudinal in nature, with minimal transverse field components.

### III. Results and Discussion

The generation of THz waves in a cylindrical plasma waveguide was investigated using Fourier-Bessel Particle-In-Cell (FBPIC) simulations and COMSOL Multiphysics for the specified configuration. FBPIC is a computational framework in cylindrical coordinates designed to study the interaction between charged particles and electromagnetic fields in relativistic regimes. This framework integrates two key components, the Particle-In-Cell (PIC) method and relativistic equations. FBPIC employs cylindrical PIC algorithms combined with an azimuthal Fourier decomposition of electromagnetic field components, making it a quasi-3D PIC code. This quasi-3D approach allows FBPIC to effectively capture three-dimensional electromagnetic field dynamics while maintaining the computational efficiency of a lower-dimensional model. The PIC method discretizes the plasma into individual particles and employs numerical techniques to solve their equations of motion. Simultaneously, the electromagnetic fields are computed on a grid and evolved according to Maxwell's equations. This framework accurately represents the complex dynamics of plasmas, including charged particle's interactions with electromagnetic fields, as well as the self-consistent evolution of plasma waves and instabilities. Additionally, COMSOL Multiphysics was employed to model the geometry of the plasma waveguide and to validate the propagation of electromagnetic waves generated by the wakefield. The numerical solutions of Maxwell's equations in COMSOL provided a complementary perspective, allowing for an examination of the boundary effects within the waveguide and ensuring consistency with the FBPIC simulations.

Figure 2 shows the variations in the electric field of the axial wakefield for different modes in the two-dimensional dielectric-plasma waveguide cross-section, as simulated using COMSOL Multiphysics. Furthermore, Figure 3 depicts the effect of various wavelengths on these variations. The results illustrate that each mode corresponds to a distinct spatial pattern of the electric field, with higher mode numbers associated with more complex field distributions. These higher modes exhibit increased oscillations and peaks within the waveguide cross-section or along the waveguide, resulting from more densely packed oscillations per unit distance and shorter wavelengths. The mode number signifies the different standing wave patterns that form within the waveguide. Higher modes are associated with shorter wavelengths. As the spatial period of oscillation decreases with an increase in mode number, more oscillations per unit distance are excited, consequently, the electric field distributions for higher modes vary more rapidly compared to the fundamental mode. The simulations also reveal that higher-order modes in the waveguide feature components that do not remain confined purely within the plasma region and extend beyond the plasma region into the surrounding dielectric. This occurs due to evanescent fields that penetrate the dielectric material, influenced by the waveguide's geometry and the dielectric's properties. This field leakage into the dielectric impacts how the wave's behavior, contributing to the dispersive nature of the waveguide system. As the wave interacts with the dielectric, different modes experience slower phase velocities due to the dielectric material's frequency-dependent response, which affects the wave's propagation. This interaction results in slower traveling waves within the waveguide, extending the distance over which the wave maintains its energy and structure before attenuating or distorting significantly.

Figure 4 demonstrates the variations of the normalized electric field of the axial wakefield behind the drive beam as a function of the normalized independent variable ($k_p\xi$) for different modes. The independent variable, *z*, represents the longitudinal distance along the waveguide. As the mode number increases, the number of oscillations and peaks within a given distance along the *z*-axis becomes more pronounced, a consequence of the shorter wavelengths associated with higher modes. Concurrently, the electric field magnitude decreases with increasing mode number.

**(a)** $n = 1$

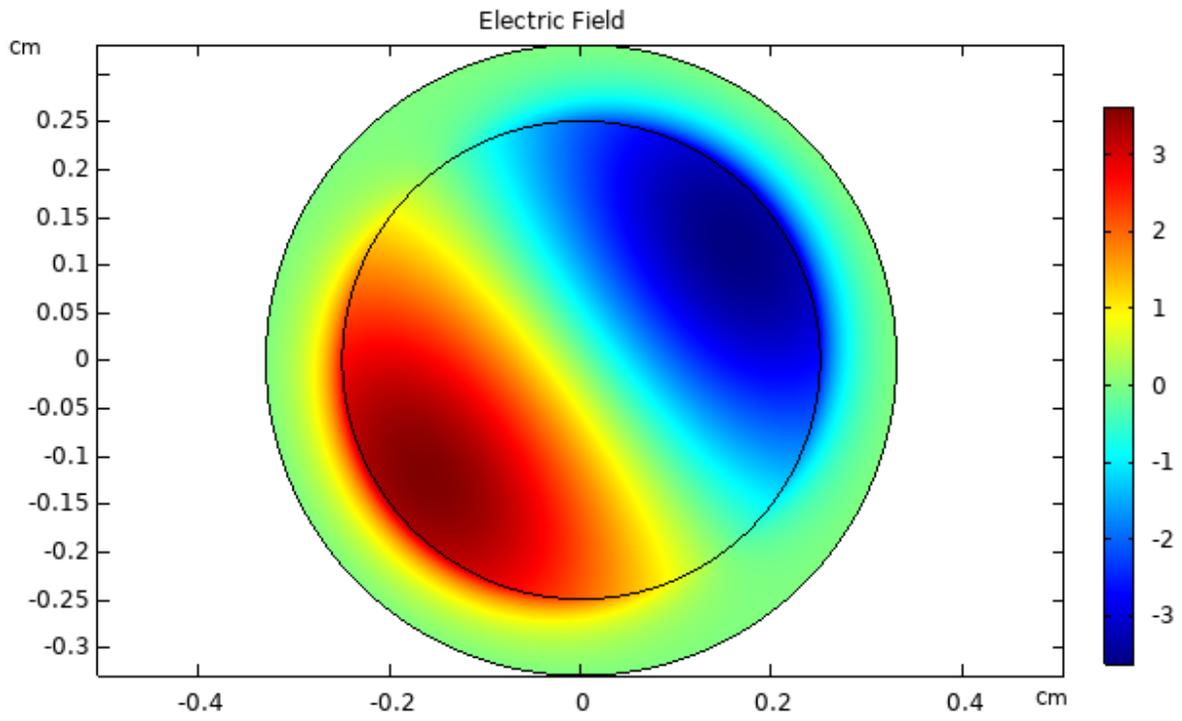

**(b)** $n = 2$

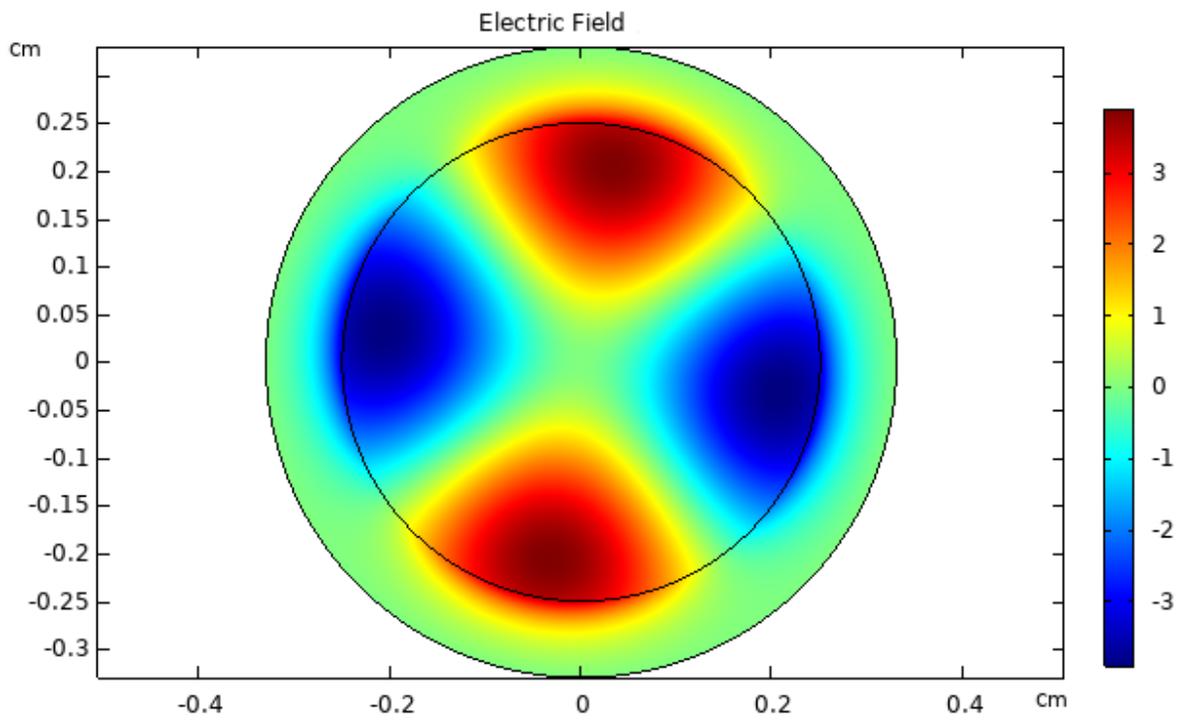

**(c)** $n = 3$

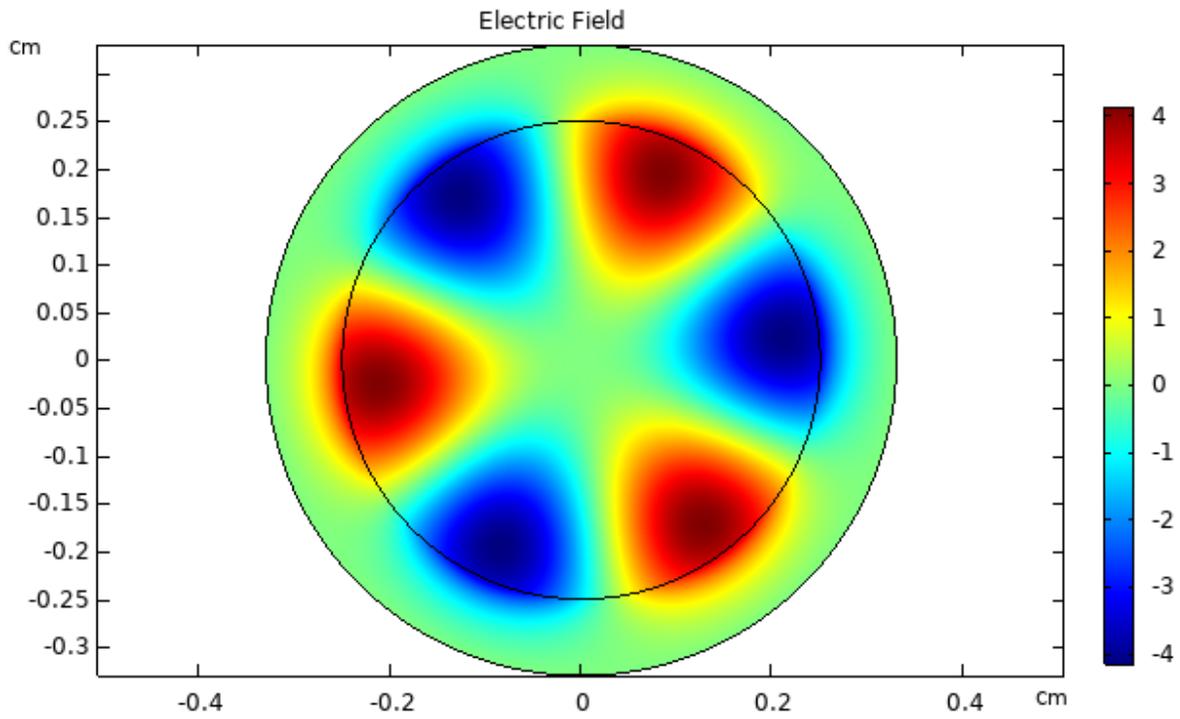

**(d)** $n = 4$

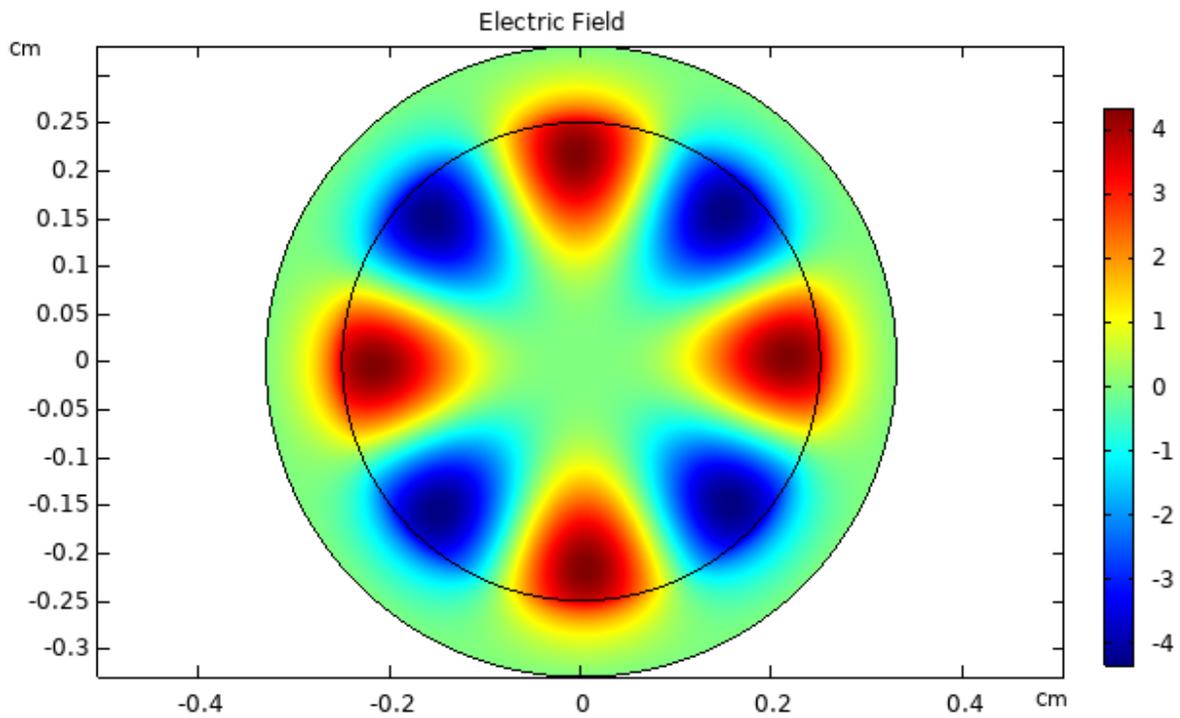

**Fig. 2.** Variations of the electric field of the axial wakefield generated behind the drive beam for different modes in the two-dimensional waveguide cross-section.

**(a)** $\lambda = 700\ nm$

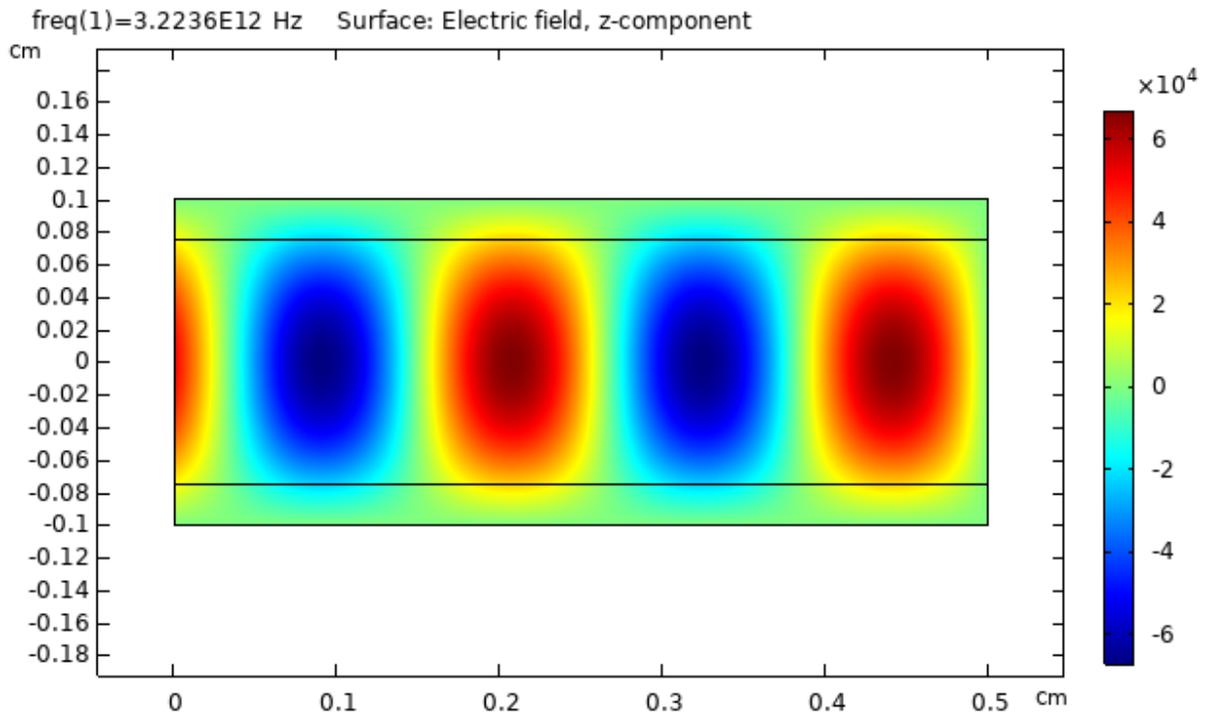

**(b)** $\lambda = 750\ nm$

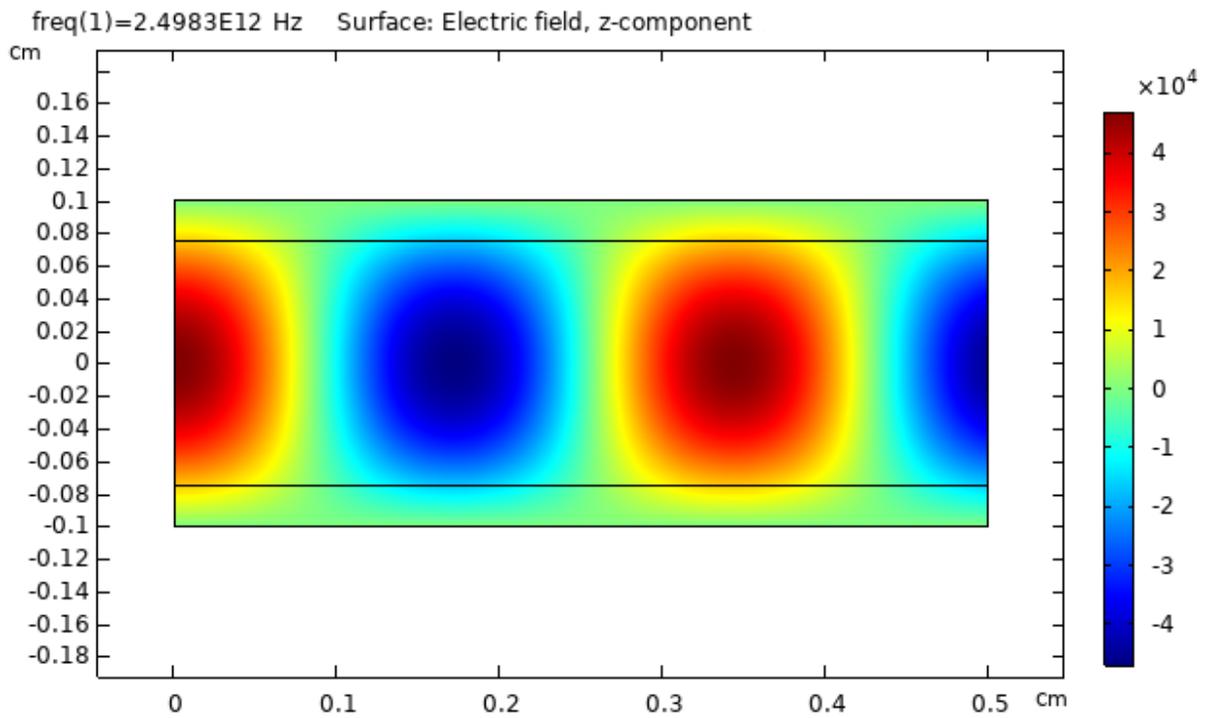

**(c)** $\lambda = 800\ nm$

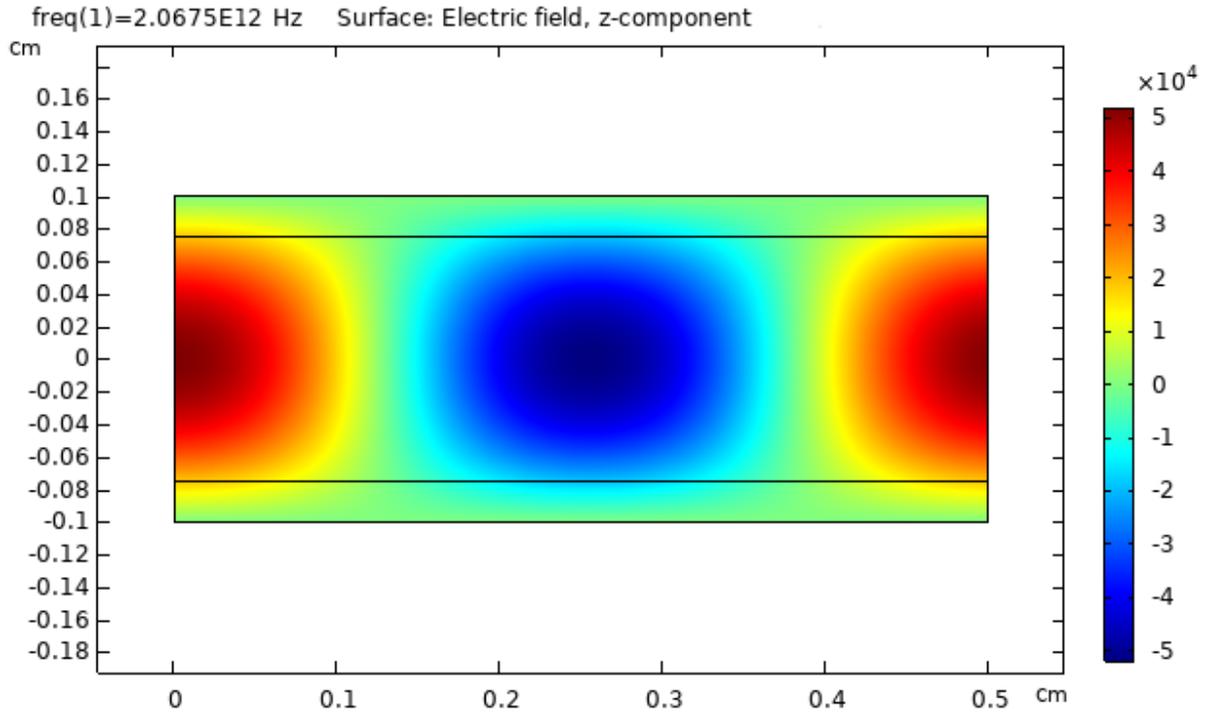

**(d)** $\lambda = 850\ nm$

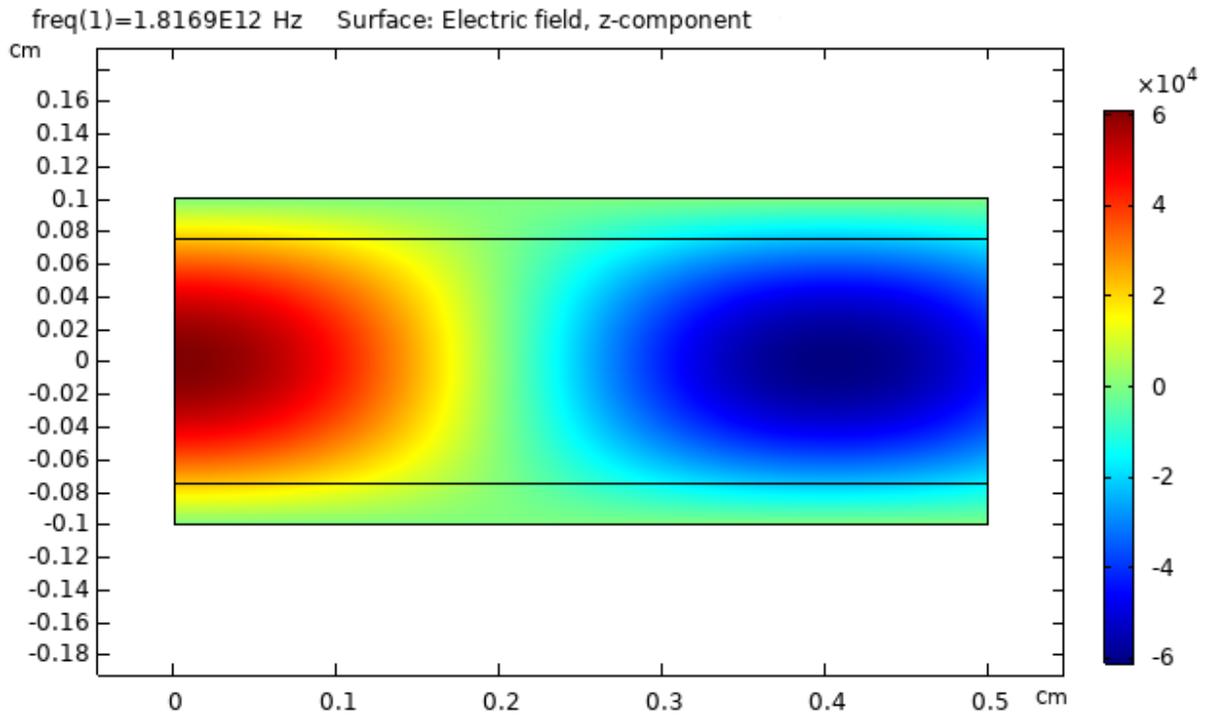

**Fig. 3.** The effect of various wavelengths on variations of the electric field of the axial wakefield generated behind the drive beam in the two-dimensional waveguide cross-section.

The modes correspond to distinct frequencies or wavenumbers, and higher mode numbers naturally lead to more oscillations within the same spatial interval, resulting in more rapid field oscillations. Since higher modes are associated with shorter wavelengths, they exhibit more peaks (oscillations) within the same range of $\xi$ with these peaks marking the points where the electric field strength reaches its maximum. In a dielectric-plasma waveguide, where wakefields are excited, lower modes typically dominate in terms of energy content, characterized by fewer oscillations and longer wavelengths. As the mode number increases, the energy is more evenly spread across a larger number of oscillations, leading to a reduction in the overall field amplitude. Furthermore, higher modes experience greater dissipation or damping, causing further reduction in the electric field magnitude as energy is lost from the system. Understanding these relationships is crucial for optimizing the design and performance of such structures in various applications.

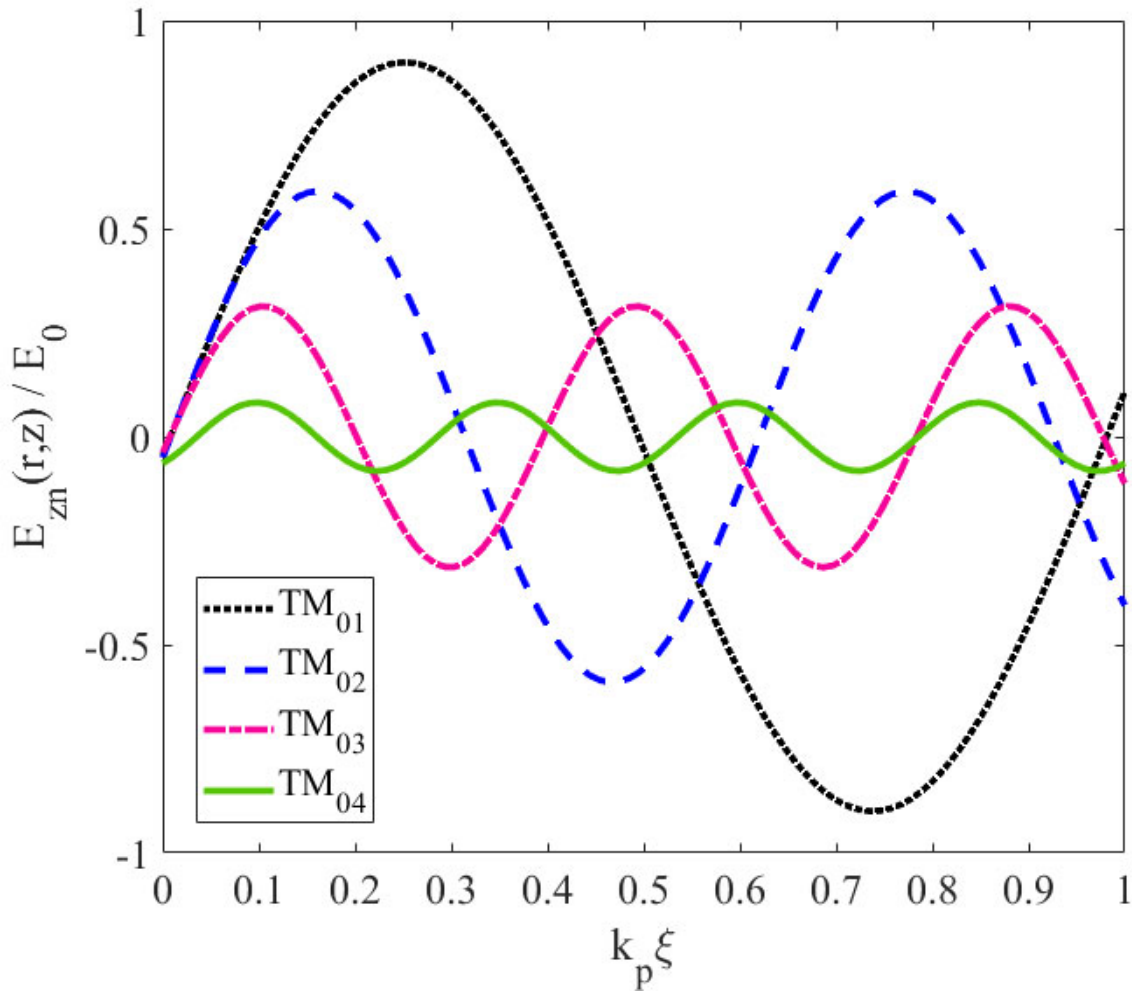

**Fig. 4.** Variations of the normalized electric field of the axial wakefield versus the normalized independent variable ($k_p\xi$) for different modes, $I_b = 2 \times 10^3 A$, $r_d = 2\ mm$, $t_d = 1\ mm$, and $L = 2\ mm$.

Figure 5 shows the impact of dielectric thickness on variations of the normalized electric field of the axial wakefield behind the drive beam, specifically for the fundamental mode. As illustrated in the figure, an increase in the dielectric thickness leads to an expansion of the outer radius of the cylindrical waveguide increases, resulting in a decrease in the electric field strength of the wakefield wave. In a cylindrical waveguide, when the dielectric thickness increases, the waveguide's outer radius expands, affecting its electromagnetic properties. In waveguides, the electric field tends to be more confined within regions of higher permittivity (i.e., the dielectric material). As the dielectric thickness increases, the effective area where the electric field can propagate becomes larger. This causes the field to spread out over a larger cross-sectional area, which in turn lowers the energy density and the electric field strength. This is analogous to how increasing the volume in which energy is distributed naturally reduces the intensity of the field at any given point. For the fundamental mode, the electric field typically has its maximum near the axis of the waveguide and decays outward. When the waveguide radius increases due to a thicker dielectric, the mode structure adjusts, leading to a more spread-out field pattern. The result is a reduction in the peak electric field amplitude, even though the total energy of the wave remains constant. This reduction is more pronounced because the fundamental mode tends to be more sensitive to changes in the waveguide geometry. Additionally, as the dielectric thickness increases, the mode tends to have a larger transverse dimension, meaning the fields are less tightly confined to the axis. As the modes spread out radially due to the larger waveguide size, the field strength decreases because the energy density is distributed over a larger radius. Furthermore, as the dielectric thickness increases, the effective refractive index seen by the electromagnetic wave also increases. This change alters the phase velocity and the wavelength of the mode. Specifically, for a given frequency, the wavelength in the dielectric-loaded waveguide becomes shorter. The reduction in phase velocity causes the electromagnetic waves to propagate more slowly, effectively shifting the mode structure and further reducing the field intensity.

Figure 6 displays the impact of pulse length on the normalized electric field of the axial wakefield behind the drive beam, plotted versus the normalized independent variable ($k_p\xi$) for the fundamental mode. As the pulse length increases, according to Eq. 16, the electric field strength decreases due to the gradual delivery of energy over time. In the dielectric-plasma waveguide system, the axial wakefield behind the drive beam is excited by an electron bunch that passes through the plasma, with the wakefield strength proportional to the charge density or the intensity of the driving pulse. A longer pulse reduces the energy density, lowering the peak electric field magnitude as energy becomes less concentrated in time. Efficient wakefield generation requires the pulse to synchronize with plasma electrons oscillations. However, longer pulses, increase the likelihood of phase mismatch between the pulse and the excited wakefield. The interference at the leading and trailing edges of the pulse leads to a destructive effect, further weakening the electric field. In addition, a longer pulse broadens the induced wakefield oscillations, reducing their sharpness peak electric field magnitude. This extended interaction diminishes the efficiency of wakefield generation, as energy and charge are less temporally concentrated, leading to a weaker overall electric field.

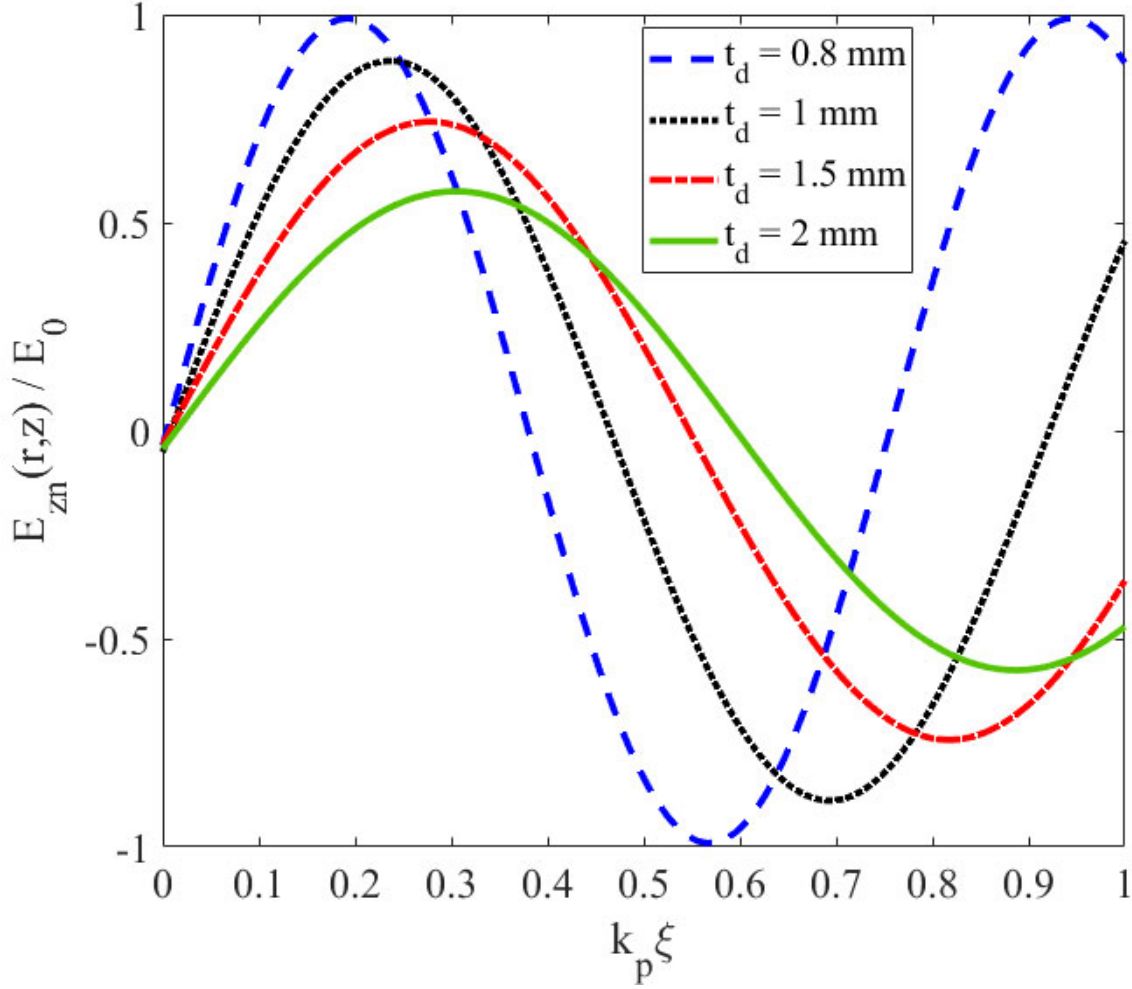

**Fig. 5.** Impact of dielectric thickness on variations of the normalized electric field of the axial wakefield versus the normalized independent variable for fundamental mode, $I_b = 2 \times 10^3 A$, $r_d = 2\ mm$, and $L = 2\ mm$.

The effect of varying total currents on the normalized electric field of the axial wakefield behind the drive beam, plotted versus the normalized variable ($k_p \xi$), is illustrated in Fig. 7. The figure demonstrates that increasing the current only enhances the electric field's magnitude of the wakefield wave without significantly altering the waveform's shape. In many wakefield generation systems, the electric field strength is proportional to the total current of the driving source (e.g., the particle beam or electromagnetic pulse). As the current increases, this proportionality leads to higher electric field magnitude while maintaining the waveform's temporal and spatial profile. A higher total current corresponds to more charge flowing through the system per unit of time. This enhanced interaction generates a substantial wakefield wave behind the drive beam. The additional energy introduced by the increased current is distributed uniformly across the waveform, leading to a stronger wakefield. Consequently, the waveform is amplified without distortion or changes to its structure. The shape and oscillatory behavior of the wakefield waveform, are primarily dictated by the characteristics of the driving pulse (e.g., bunch shape, frequency components), as well as the properties of the medium in which the wakefield is generated. These factors remain constant regardless of changes in current, ensuring the waveform's shape remains consistent. Therefore, increasing current results solely in a stronger electric field while preserving the waveform's profile.

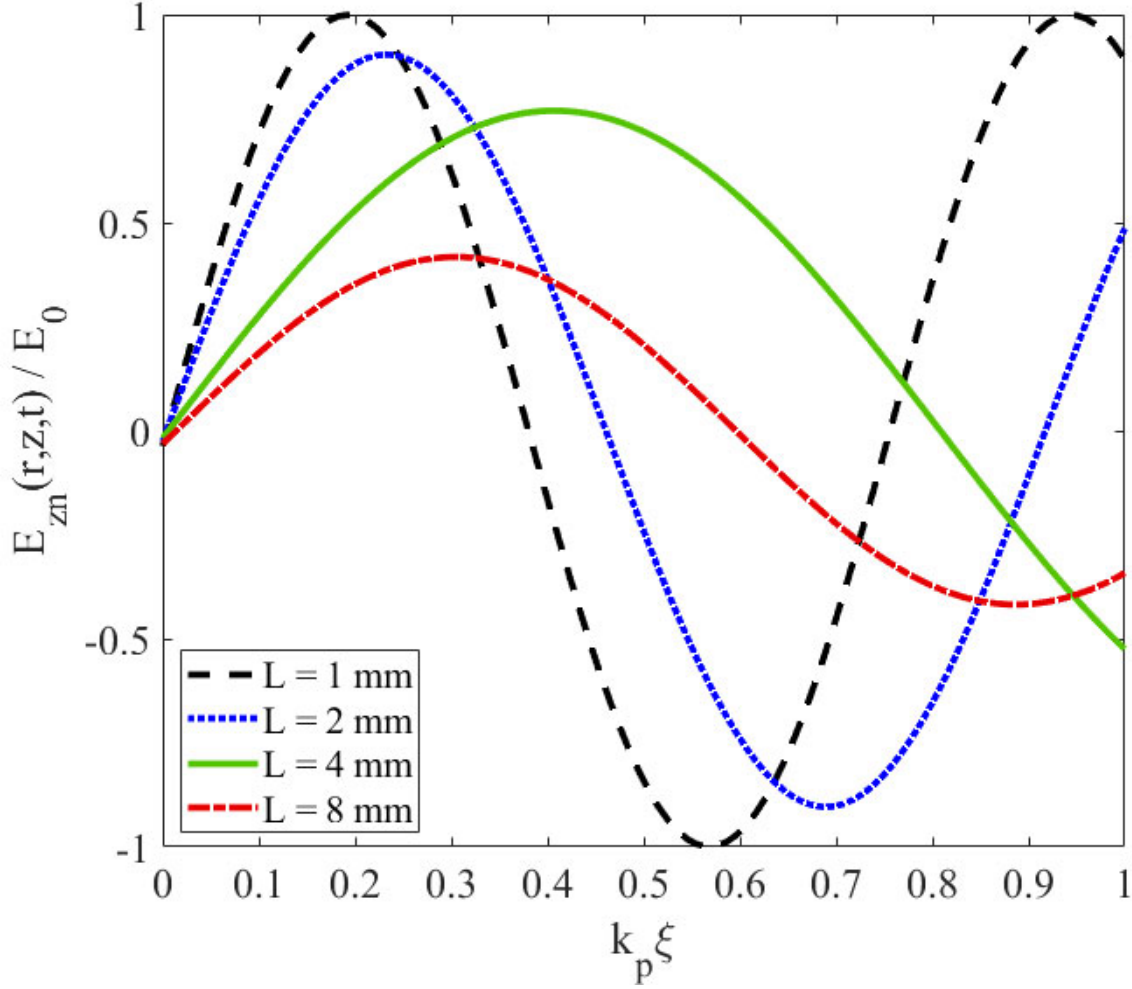

**Fig. 6.** The role of pulse length on variations of the normalized electric field of the axial wakefield versus the normalized independent variable for fundamental mode, $I_b = 2 \times 10^3 A$, $r_d = 2\ mm$, and $t_d = 1\ mm$.

Figure 8 highlights the role of the driver beam radius on the normalized electric field of the axial wakefield behind the drive beam, plotted versus the normalized independent variable ($k_p\xi$) for fundamental under varying parameters. The figure demonstrates that as the driver beam radius increases, the magnitude of the electric field of the wakefield decreases. Additionally, beyond a certain radius, the direction of the electric field oscillations reverses. This behavior indicates that the driver beam radius plays a crucial role in shaping the distribution of charge or energy that excites the wakefield. As the beam radius increases, the charge or energy spreads over a larger cross-sectional area. This reduces the beam's charge density (charge per unit area). Since the wakefield strength depends on charge density, a larger beam radius leads to a weaker excitation and reduced electric field magnitude. The reversal in the electric field's oscillation direction at larger radii can be attributed to mode interaction and phase shifts in the wakefield dynamics. Wakefields are described by a superposition of modes, including the fundamental mode and potentially higher modes, whose phase and amplitude depend on the beam radius. As the radius grows, the coupling between the driver beam and these modes changes, altering their phase relationships. The weaker interaction occurs because the driver beam is no longer optimally coupled to the plasma electrons in the same manner as it was with a smaller radius. The beam's increased size causes destructive interference between modes, resulting in a 180-degree phase shift and the observed field direction reversal. Furthermore, in a plasma wakefield system, the driver beam excites plasma oscillations that generate the wakefield. The plasma oscillations are influenced by the plasma density and the beam's spatial distribution. The plasma wavelength, linked to the plasma density, determines the wakefield's resonance with plasma oscillations. When the beam radius increases significantly, it no longer efficiently drives the original fundamental mode, causing a shifting of the system's resonant behavior. The plasma coupling potentially alters the wakefield, contributing to the observed reversal in the electric field oscillations.

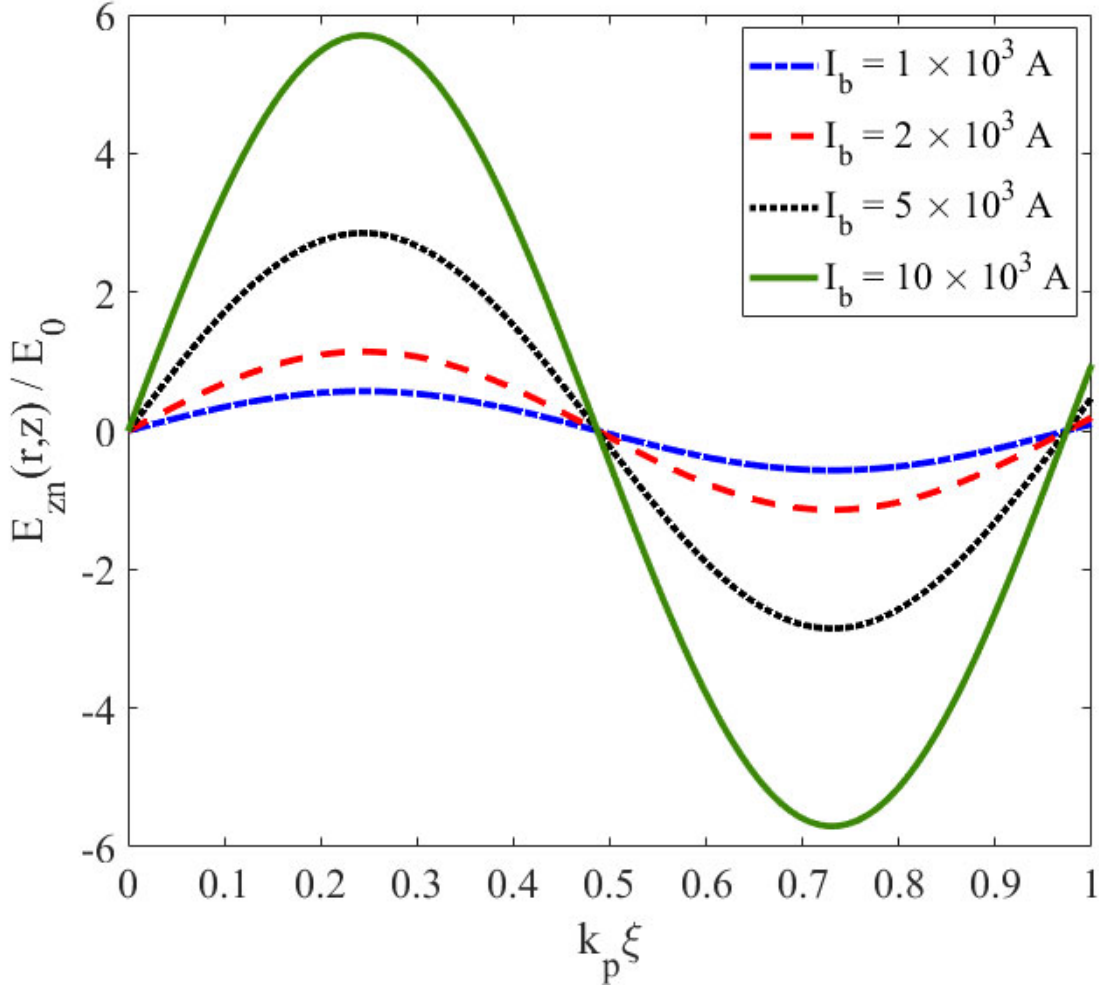

**Fig. 7.** Variations of the normalized electric field of the axial wakefield versus the normalized independent variable for various total current and for fundamental mode, $r_d = 2\ mm$, $t_d = 1\ mm$, and $L = 2\ mm$.

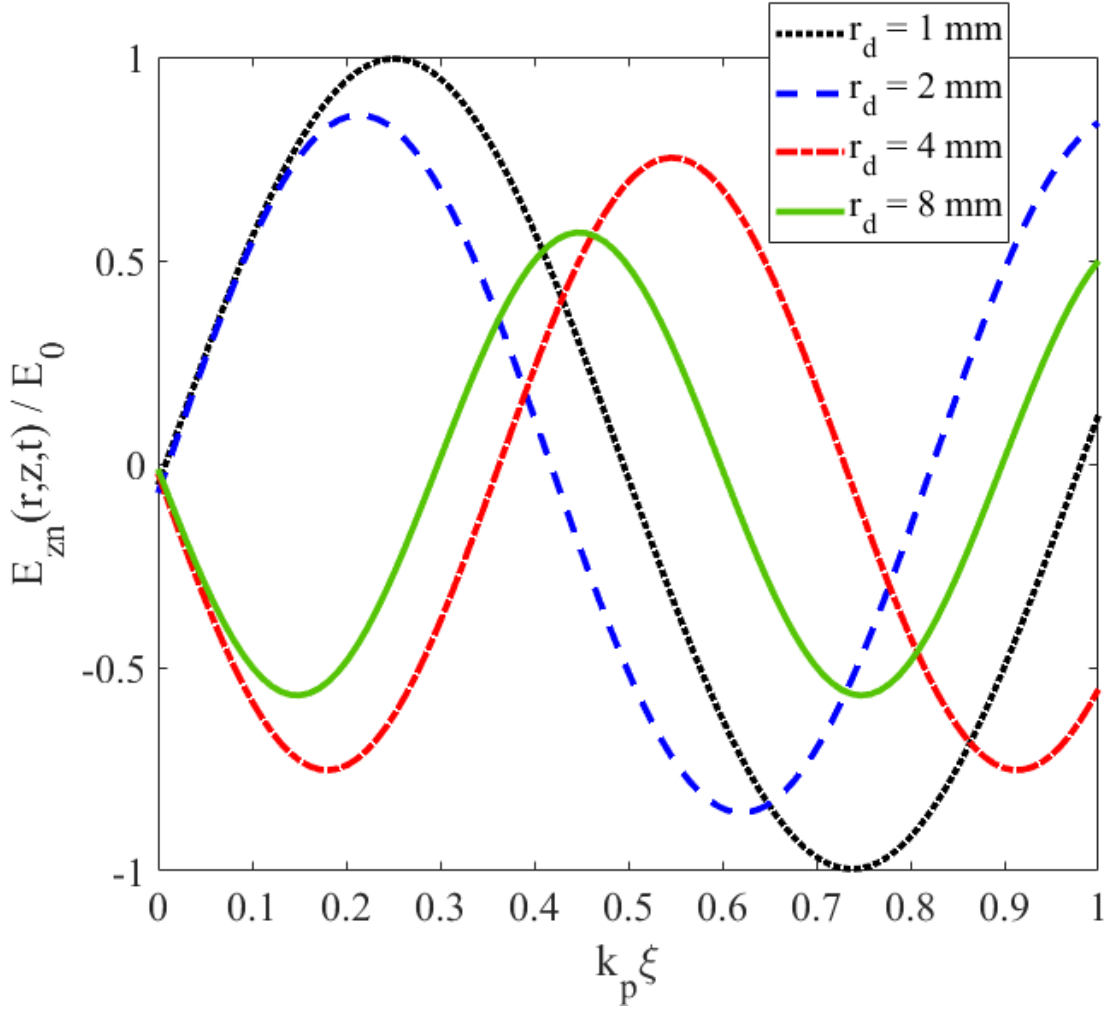

**Fig. 8.** The effect of the driver beam radius on variations of the normalized electric field of the axial wakefield versus the normalized independent variable for fundamental mode, $I_b = 2 \times 10^3 A$, $t_d = 1\ mm$, and $L = 2\ mm$.

Figure 9 illustrates the impact of the various dielectrics on the normalized power of the axial wakefield behind the drive beam as a function of the normalized independent variable ($k_p \xi$) for the fundamental mode. The figure shows that as the dielectric constant increases, the normalized power of the wakefield also rises significantly. Power is a measure of energy transfer per unit of time, depending on how effectively the energy of the driving beam is converted into the wakefield's electromagnetic energy. Since power is proportional to the square of the electric field, even modest increases in the electric field due to a higher dielectric constant can substantially amplify the normalized power. A higher dielectric constant indicates a more polarizable material, enabling it to support stronger electromagnetic fields for the same applied energy. This allows the material to store more electric energy in the form of displacement fields, enhancing the waveguide's ability to sustain stronger electromagnetic fields and increasing the wakefield's power. In a dielectric-plasma waveguide, the dielectric constant also influences the boundary conditions, affecting the wavelength, phase velocity, and mode structure of the wakefield A higher dielectric constants confine the electric field more strongly within the dielectric material, improving the coupling of the drive beam's energy to plasma electrons and increasing energy transfer efficiency. Moreover, a higher dielectric constant reduces the phase velocity of the wakefield within the dielectric material. This reduction enables greater energy deposition from the drive beam into the wakefield, further amplifying the power.

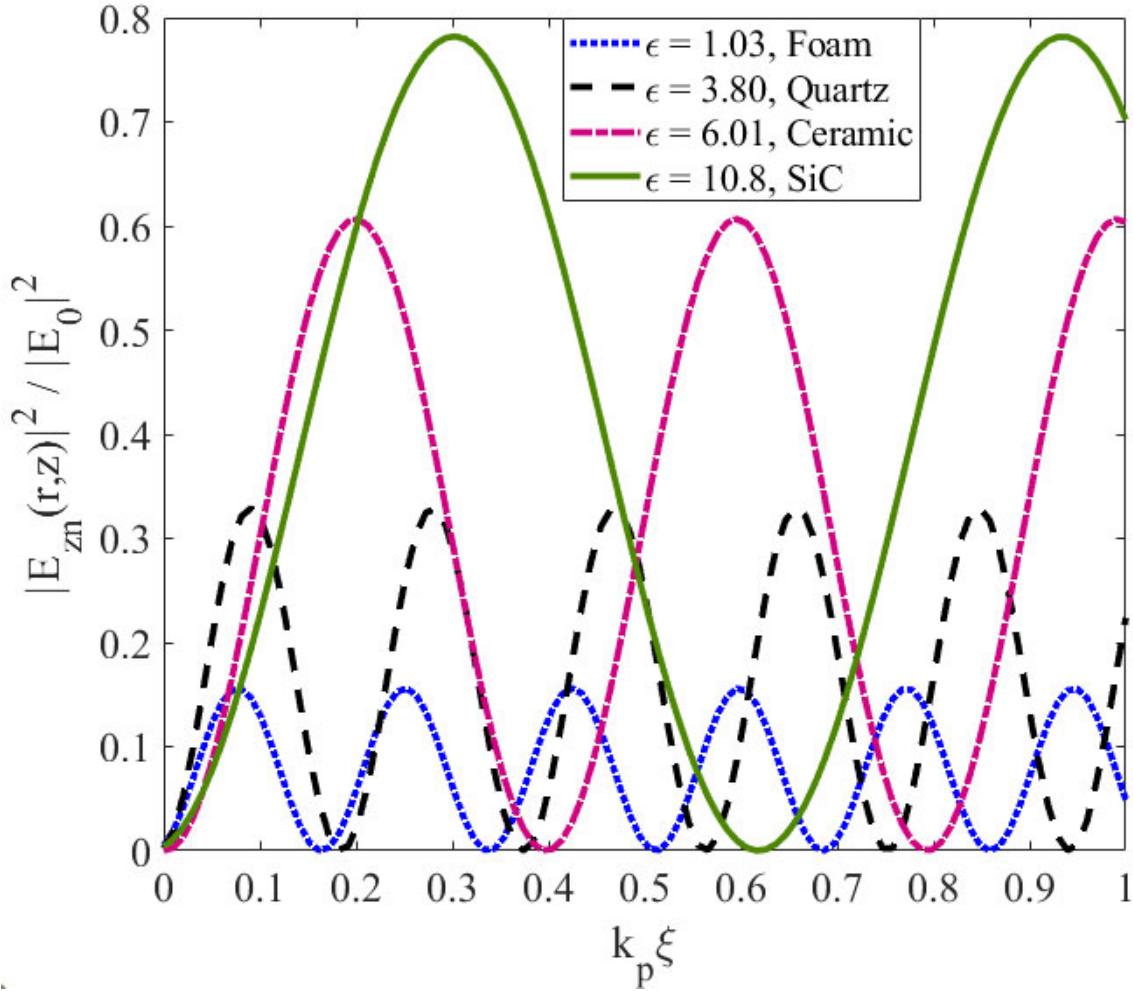

**Fig. 9.** Variations of the normalized power of the axial wakefield versus the normalized independent variable for various dielectrics for fundamental mode, $I_b = 2 \times 10^3 A$, $r_d = 2\ mm$, $t_d = 1\ mm$, and $L = 2\ mm$.

## IV. Conclusions

The generation of THz radiation through wakefield excitation in a cylindrical dielectric plasma structure, employing a hybrid approach that integrates dielectric and plasma wakefield is explored. This combination leverages the strengths of both materials: dielectrics efficiently transfer energy to high-energy particles, while plasmas sustain intense electric fields, leading to efficient wakefield excitation, yielding higher-energy particles and more intense THz radiation. The results of the present study reveal that a high-energy laser pulse propagating through a plasma-loaded dielectric waveguide induces wakefields that drive the emission of THz radiation. This dual capability, enhancing both gradients and THz wave output, highlights the versatility and potential of the proposed structure. The advanced numerical simulations complement the theoretical analysis. The FBPIC code is specifically optimized for a cylindrical coordinate system, and effectively models the cylindrical plasma waveguide. The code by incorporating relativistic equations, FBPIC accurately models high-energy particle dynamics and their interactions with electromagnetic fields. The electromagnetic fields are decomposed into azimuthal Fourier components, reducing the computational complexity while preserving essential three-dimensional dynamics. To validate the results, COMSOL Multiphysics was utilized to model the plasma waveguide geometry and analyze the propagation of the generated electromagnetic waves. By solving Maxwell's equations, COMSOL provided insight into boundary effects and confirmed consistency with the FBPIC simulations. Additionally, the study systematically examined the effects of key parameters including dielectric thickness, DC external magnetic field, pulse length, driver beam radius, and total current on the wakefield generation and THz radiation. The variation in these parameters reveals their influence on wakefield characteristics and the resulting THz wave radiation. This comprehensive analysis provides valuable insights for optimizing the dielectric-plasma wakefield structure, enhancing both the efficiency and quality of THz wave generation.

**Acknowledgment**

This work is based on research funded by Iran National Science Foundation (INSF) under project No. 4021340. The authors would like to express their gratitude towards INSF.

**Author Contributions**

Ali Asghar Molavi Choobini was responsible for conceptualization, data curation, formal analysis, investigation, methodology, writing the original draft, and review and editing, all with equal contributions.

Farzin M. Aghamir contributed equally to investigation, project administration, supervision, validation, and review and editing of the manuscript.

**Data availability statement**

The data that support the findings of this study are available from the corresponding author upon reasonable request.

**Competing interests**

The authors declare no competing interests.